\documentclass[epj]{svjour}
\usepackage{epsf,psfig}
\newcommand \beq{\begin{eqnarray}}
\newcommand \eeq{\end{eqnarray}}
\newcommand{\set}[2]{\newcommand{#1}{#2}}
\set{\pa}{\partial \over \partial\, }
\set{\leftvector}{\stackrel{\leftarrow}{\partial }}
\set{\rightvector}{\stackrel{\rightarrow}{\partial }}
\begin{document}
\title{Formation of correlations and
energy-conservation at short time scales}
\author{K. Morawetz\inst{1} \and H. S. K\"ohler\inst{2}
}                     
%
%
\institute{Fachbereich Physik, Universit\"at Rostock,
18051 Rostock, Germany\and Physics Department,
University of Arizona,Tucson,Arizona 85721}
\date{Received: date / Revised version: date}
%
\abstract{
The formation of correlations due to collisions in an
interacting nucleonic system is investigated.
Results from one-time kinetic equations are compared with the
Kadanoff and Baym two-time equation with collisions included in
Born approximation.
A reasonable agreement is found for a proposed
approximation of the memory effects by a finite duration of
collisions. This form of collision integral is in
agreement with intuitive estimates from Fermi's golden
rule. The formation of correlations and the build up time is calculated analytically for the high
temperature and the low temperature limit. Different approximate expressions are compared with the numerical results. We
present analytically the
time dependent interaction energy and the formation time for Gau\ss{}-
and Yukawa type of potentials.
} 
\maketitle

\section{Introduction}

The Boltzmann transport equation has played a very important role in the
development of non-equilibrium statistical mechanics. This microscopic
equation describes the time-evolution of a distribution-function in
phase-space and has also provided a connection with macroscopic hydrodynamic
equations by a moment expansion of the momentum. Important applications
are for example
the well-known Chapman-Enskog calculations of transport coefficients.
In later developments the Markovian Boltzmann-equation has been extended
to include memory and correlation-effects in the collision-integral
and there are a large number
of publications concerning such improvements. These classical kinetic
equations describe the time-evolution of a one-time distribution
function $f({\bf r,p},t)$.

Meanwhile, a quantum two-time theory for the
time-evolution of real time Green's functions $G({\bf r,p},t,t')$ has been
developed using the Schwinger-Keldysh formalism. The quantum image of the
classical Boltzmann equation is usually referred to as the Kadanoff-Baym
(KB)
equations \cite{KB62}. These equations have often been considered too
complicated to solve numerically in the past. However, several numerical applications exist now. The Kadanoff-Baym
equations have also played an important
role in the improvements of the Boltzmann equation especially by using
the \it Generalised Kadanoff Baym Ansatz \rm (GKB) of Lipavsky et al \cite{LSV86}. This
ansatz allows a reduction of the two-time formalism to a formally
simpler one-time formalism e.g. the Boltzmann equation.
The time off-diagonal Green's function
elements are related by GKB to the time-diagonal by the spectral-functions.
By various approximations of the spectral-functions various one-time
approximations of the two-time equations can be obtained.(See e.g.
\cite{hsk96})

These kinetic equations describe different relaxation stages.
During the very fast first stage, correlations imposed by the initial
preparation of the system
are decaying \cite{B46,BKSBKK96}. These are contained in
off-shell or dephasing processes described by two-time propagators. During this stage of
relaxation the quasiparticle picture is established \cite{LKKW91,MSL97a}.
After this very fast process the second state develops
during which the one-particle distribution relaxes towards the equilibrium
value \cite{RTb90} with a relaxation time $\tau_{\rm rel}$.
First the momentum anisotropy relaxes by small angle scattering events and
then the energetic degrees of freedom relax. During this
relaxation state the virial corrections are established and can
be consistently described by a nonlocal Boltzmann kinetic
equation \cite{SLM96,LSM97}.
The time of the first stage $\tau_c$ is mostly shorter than the relaxation
time $\tau_{\rm rel}$ of one particle distributions which is entirely
determined by the collision process. We will focus on the first stage
which is related to the formation of correlations.

The formation of correlations is connected with an increase of the kinetic energy or equivalently the build
up of correlation energy. This is due to rearrangement processes which let decay higher order correlation
functions until only the one - particle distribution function relaxes. Because the correlation energy is a
two - particle observable we expect that the relaxation of higher order correlations can be observed best
within this quantity. Of course, the total energy of the system is conserved
\begin{eqnarray}
\frac{\partial}{\partial t} \left ( \langle \frac{p_1^2}{2 m} \rangle (t) +
   E_{\rm corr} (t)\right ) &=& 0\label{cons},
\end{eqnarray}
which means that the kinetic energy increases on cost of the correlation energy $E_{\rm corr}(t)$.
We will observe a transformation of correlation into kinetic energy. This process saturates on the end of the
first stage of relaxation. It is more convenient to calculate the kinetic energy than the correlation energy
because
the kinetic energy is a one-particle observable. Consequently, the time dependence of the kinetic energy will
be investigated within the kinetic theory. This can only be accomplished if we employ a kinetic equation which
leads to the total energy conservation (\ref{cons}). It is immediately obvious that the ordinary Boltzmann
equation cannot be appropriate for this purpose because the kinetic energy is in this case an invariant of the collision
integral and constant in time. Imposing the conservation of the form (\ref{cons}) we have to consider
non-Markovian kinetic equations \cite{M94}, which account for the formation of two particle correlations.

Within these kinetic equations the collision integral is an
expression of the two-particle correlations. While the one-particle distribution remains almost unchanged
during the first stage of relaxation, the two-particle correlations relaxes. Consequently the one- particle
spectral function is changing. The latter one is responsible for the dephasing and therefore formation of
correlations.
Analytical expressions for the time dependence of the kinetic and correlation energy are obtained in this paper
by considering explicitly this dephasing process.

We start from a kinetic equation appropriate for short time scale in Born
approximation.
It contains the full
memory-effect but no damping i.e. no explicit width
of the spectral function, because quasiparticles are not
yet formed on this time scales.
In Chapter II we give an overview of the gradient approximation
with emphasis on energy-conservation and correlation energy.
In appendix \ref{append} we discuss the limit of
complete collisions and the weakening of initial correlations.
In appendix \ref{appc} we calculate the equilibrium value of
the correlation energy for high and low temperature
limits analytically using Gaussian and Yukawa type interactions.

On the very first time scale we can neglect retardation effects in the one
particle distribution function, but we have to
keep into account off-shell properties of the collision integral.
Therefore we use the {\it finite duration } approximation in
Chapter III.
It leads to the correct equilibrium value and is intuitively
clear from Fermi's Golden rule. It is compared numerically
with the KB results.
These calculations also bring to the attention the correlation-time i.e.
the time for build-up of correlations.

From the observation that the time-variation of the
distribution-functions can be neglected in the first stage of relaxation we
obtain an
analytic expression for the time dependent formation of correlations.
Especially we give analytical results for the formation
time of correlations in a high and in a low temperature limit.
Comparisons are made with numerical calculations.

Chapter IV summarizes our results and we discuss some
aspects regarding the correlation time.
The appendices show some important relations necessary for our analytic
calculations.

\section{Correlation Energy in gradient expansion}
The kinetic equation in Born approximation for spatial
homogeneous media
including complete time convolution (memory effect) but no damping
is called Levinson equation and reads
\cite{L65,L69,JW84,MWR93}
\begin{eqnarray}\label{kinetic}
&&   \frac{\partial}{\partial t} f(p_1)
     =\nonumber\\
&& \frac{2 s_1 s_2}{\hbar ^2}
     \int \frac{dp_2 dp_1' dp_2'}{(2 \pi \hbar )^6}
     V(\mid p_1 - p_1'\mid )^2
     \delta (p_1 + p_2 - p_1' - p_2')
     \nonumber \\
  &\times& \int_{t_0}^{t }d\tau {\rm cos}
(\frac{1}{\hbar} (E_1 + E_2 - E_1' - E_2')(t-\tau)
)\nonumber\\
  &\times &
(f(p_1', \tau ) f(p_2', \tau )
{\bar f}(p_1, \tau )
{\bar f}(p_2, \tau ) \nonumber\\
&-& f(p_1, \tau ) f(p_2, \tau ) {\bar f}(p_1',
\tau )
{\bar f}(p_2', \tau ) )
     \nonumber \\
\end{eqnarray}
with $\bar f = 1-f$, the free particle dispersion
$E=p^2/2m$ and the spin-isospin degeneracy $s_1,s_2$.
The distribution functions are normalized to the density as $s \int {d p\over (2 \pi \hbar)^3} f(p)=n$. For the sake of simplicity we have omitted the Hartree and Fock contribution. Since we discuss the correlation energy the meanfield contribution is just additive. The dispersion $E(p)$ in the collision integral is modified by the Fock term, but we use this effect only in an approximative way by understanding $m$ as effective mass.

The Boltzmann collision integral is obtained from
equation (\ref{kinetic})
if: (i) One neglects the time
retardation in
the distribution functions, i.e. the memory effects
and (ii) The finite initial time $t_0$ is set equal to
$-\infty$ corresponding to what is usually referred to as
the limit of complete
collisions.
The memory effect is condensed in the explicit retardation of the distribution function. This would lead to
gradient
contributions to the kinetic equation which can be shown to be responsible for the formation of high energetic
tails in the distribution function \cite{MR95,SL95}. This effect will be established on the second stage of
relaxation.

The second effect is contained in the energy broadening or off-shell behavior in (\ref{kinetic}). This is
exclusively related to the spectral properties of the one-particle propagator and therefore determined by
the relaxation of two-particle correlation.
Since we are studying the very short time region after the initial disturbance we can separate the one-particle
and two-particle relaxation. On this time scale the memory in the distribution functions can be neglected but
we will keep the spectral relaxation implicit in the off-shell $\cos$-function of (\ref{kinetic}).
This effect is the most relevant one
for obtaining
the time evolution of the interaction (or correlation) energy and
therefore
energy conservation.

In the following discussion we shall only be concerned with the
time integration.
Therefore we introduce the short hand notation of
equation (\ref{kinetic})
\begin{equation}\label{short}
 \frac{\partial}{\partial t} f(p_1)=\frac{ 1}{ \hbar^2}
\int_{t_0}^{t }d\tau
\cos{{\Delta E (t-\tau ) \over \hbar}} \; F(\tau),
\end{equation}
where
\beq
  F(\tau)&=&
   2 V(\mid p_1 - p_1'\mid )^2
     \delta (p_1 + p_2 - p_1' - p_2')  \nonumber\\
 && \times
(f(p_1', \tau ) f(p_2', \tau ) {\bar f}(p_1, \tau )
{\bar f}(p_2, \tau ) \nonumber \\
&&- f(p_1, \tau ) f(p_2, \tau ) {\bar f}(p_1',
\tau ) {\bar f}(p_2', \tau ) )\nonumber \\
\label{short2}
\eeq
and the 9-dimensional momentum-integration
is suppressed in
Eq. (\ref{short}) and in the following.

From this equation one derives
balance equations
by integration over momentum $p_{1}$. The first two moments,
i.e. the density and
total linear momentum, are
conserved. For the Markovian Boltzmann equation the kinetic energy is
conserved, while potential energy is zero. In the present case
including the memory effect one finds \cite{M94}
\begin{eqnarray}\label{energy}
\frac{\partial}{\partial t} ( \langle \frac{p_1^2}{2 m}
\rangle +
 E_{\rm corr}) &=& 0,
\end{eqnarray}
where the correlation energy $E_{\rm corr}$
is given by
\begin{eqnarray}\label{energ}
E_{\rm corr}(t)-E_{\rm corr}(t_0) &=& -\frac{1}{4 \hbar }
\langle \int_{0}^{t-t_0 }\!\!d\tau \, \sin{ \left({\Delta E\tau
 \over \hbar}\right )} F(t-\tau) \rangle.\nonumber\\
&&
\end{eqnarray}
Here $<>$ indicates the integration over $p_1$. We like to point out that we have neglected initial correlations in the kinetic equation (\ref{kinetic}) in agreement with the studied sudden switching approximation. Consequently, $E_{\rm corr}(t_0)$ describes only possible constant background correlations not formed by the binary collisions.
Expanding $F(t-\tau)$ around $t$ one obtains a
gradient expansion series for the interaction energy that reads
\beq
E_{\rm corr}(t)-E_{\rm corr}(0)&=&\sum\limits_{n=0}^{\infty} <V_n(t)
F^{(n)}(t)>\nonumber\\
V_n(t)&=&- \frac {1}{ 4 \hbar} {(-1)^n \over n !}
\int\limits_0^{t-t_0} d t' t'^n \sin{{ \Delta E t'
\over \hbar}},\nonumber\\\label{s1}
\eeq
where the n-th time
derivative of $F(t)$ is given by
$F^{(n)}(t)={\partial^n \over \partial t^n} F(t)$.

Taking the time-derivative of (\ref{s1}) one finds
\beq
{\pa t} E_{\rm corr}&=&\sum\limits_{n=0}^{\infty}
<V'_n(t)F^{(n)}(t)+V_n(t)F^{(n+1)}(t)>.\nonumber
\\\label{s3}
\eeq

On the other hand one
can express the time derivative of the
interaction energy in terms of a
gradient expansion of the collision integral directly from Eq.
(\ref{kinetic}).
This leads to
\beq
{\pa t} E_{\rm corr}&=&\sum\limits_{n=0}^{\infty} <I_n(t)
F^{(n)}(t)>\nonumber\\
I_n(t)&=&- \frac {1}{ 4 \hbar} {(-1)^n \over n !}
\int\limits_0^{t-t_0} d t' t'^n \Delta E \cos{{ \Delta
E t' \over \hbar}}.\nonumber\\\label{s2}
\eeq
Note the difference between the two expansions
(\ref{s2}) and (\ref{s3}). For example, the zero order
term in Eq.(\ref{s3}) does contain not only the zero order term but also
part
of the first order term of Eq.(\ref{s2}).
This is understandable, because the collision integral
determines the \it time derivative \rm (\ref{s2}) of the correlation
energy. One
has to expand
the collision integral
one step further
in order
to obtain the correlation energy (\ref{s1}) up to a specific level
of
gradient expansion. This is a quite general observation
for any order of gradient approximation.

Comparing the two gradient expansions (\ref{s2}) and (\ref{s3})
we establish a relation between
$I_{n}(t)$ and $V_{n}(t)$
\beq
I_n&=&{\pa t}V_n+V_{n-1}\nonumber\\
I_0&=&{\pa t} V_0
=-\frac {1}{ 4 \hbar} \sin{{ \Delta E (t-t_0)\over
\hbar}},
\label{z1}
\eeq
or inversely
\beq
V_n(t)&=&\int\limits_{t_0}^t dt'
(I_n(t')-V_{n-1}(t'))\nonumber\\
V_0(t)&=&\int\limits_{t_0}^t dt' I_0(t')=\frac {1}{ 4 } {\cos{{ \Delta E (t-t_0)\over
\hbar}}-1 \over \Delta E}.
\label{z2}
\eeq
The long time limit of the different gradient approximations of the kinetic equation are presented in appendix \ref{append} and is found to be unique.
The limit of complete collisions $t_0\rightarrow -\infty$ and the connected problem of weakening of initial correlations are discussed there.

\section{The formation of correlations}


To lowest order
the gradient expansion (\ref{s1}), the correlation energy is
\beq\label{vv}
E_{\rm corr}(t)-E_{\rm corr}(0) &=& \frac 1 4 \langle
\frac{\cos{{\Delta E
(t-t_0)\over \hbar}}-1}{\Delta E}
F^{(0)}(t) \rangle. \nonumber\\
\eeq

Retaining only the first term for this correlation energy is
equivalent to
an approximation of
the non-Markovian collision integral (\ref{kinetic})
where we neglect the time dependence of the
distribution functions while keeping the finite initial time
$t_0$. This approximation gives instead of (\ref{short})
\beq\label{short1}
 &&\frac{\partial}{\partial t} f(p_1)=\frac 1 \hbar
{\sin{{\Delta E (t-t_0 ) \over \hbar}} \over \Delta E
}\; F(t)\nonumber\\
&=& \frac{2}{\hbar }
     \int \frac{dp_2 dp_1' dp_2'}{(2 \pi \hbar )^6}
     V(\mid p_1 - p_1'\mid )^2
     \delta (p_1 + p_2 - p_1' - p_2')
     \nonumber \\
  &\times& {\sin{(E_1 + E_2 - E_1' - E_2')(t-t_0)
/\hbar} \over E_1 + E_2 - E_1' - E_2'}\nonumber\\
  &\times &
[f(p_1', t ) f(p_2', t )
{\bar f}(p_1, t )
{\bar f}(p_2, t ) \nonumber\\
&-& f(p_1, t ) f(p_2, t ) {\bar f}(p_1',
t )
{\bar f}(p_2', t ) ].
     \nonumber \\
\eeq
Using the same steps which was used
to derive the correlation energy
$E_{\rm corr}(t)$ in
(\ref{energ})  from the collision integral
(\ref{kinetic}) one easily finds that the collision integral
(\ref{short1}) gives the lowest order term
of the time derivative of this correlation energy, i.e. (\ref{vv}).
The off-shell function in this collision integral contains a memory of the initial state at $t_0$. This induces a
memory in the kinetic equation in spite of the fact that the collision integral is formally Markovian.
From (\ref{s1}) one recognizes that the equilibrium or long time
limit of the correlation energy is exactly give by the first
order gradient approximation (\ref{vv}) since all next orders
include time derivatives and vanishes therefore on the long time scale.

We summarize three important features of this
approximate collision integral: (i) It reproduces the zero order
gradient term of the time dependent correlation energy (\ref{s1}).
(ii) It leads to the complete expression for the
equilibrium correlation energy (\ref{eq}) and (iii) it
is a direct consequence of Fermi's Golden Rule, where
we keep time dependent perturbation theory.
Previously, this approximation has been considered also
for electron plasmas \cite{BKSBKK96,MSL97a,KBBS96,BK96}.

It was shown in appendix \ref{append} that the two operations of taking the
time-derivative and the long-time limit of $t_{0}$ do not commute.
This is nicely illustrated by Eq. (\ref{short1})
which gives
the correct correlation
energy (\ref{eq}) if the limit $t_0\rightarrow -\infty$ is
performed afterwards.
If however the limit of complete
collision $t_0 \rightarrow -\infty$ (or $t \rightarrow
\infty$) is performed first on the kinetic equation
(\ref{short1}) it reduces to the Markovian Boltzmann equation without
correlation energy.

We refer to Eq.  (\ref{short1}) as the {\it finite duration }
time approximation. It carries the most important features of the build up of
correlations after the interactions are switched on in the initially uncorrelated system.
This will be demonstrated by some numerical examples below.
We shall first present some analytical results for high and low
temperature limit and compare
them with the numerical solution later.
To this end we assume a system consisting of two different types
of particles $a,b$ with
different masses $m_a,m_b$. As a first example
we shall consider a Yukawa-type potential of the form
\beq
V_{\rm Y}(r)={g_{ab} \over r} {\rm e}^{-\kappa r}
\label{pot}
\eeq
where in nuclear physics applications $g_{ab}$ is the
coupling and $\kappa$ the effective range of potential
given by the inverse mass of interchanging mesons. In
plasma physics applications the potential (\ref{pot})
represents the Debye potential with $g_{ab}=e_a e_b/\epsilon_0$
given by the two charges and the inverse screening
length
\beq
\kappa^2=\sum\limits_a {4 \pi e_a^2\over \epsilon_0} {\partial n_a \over \mu_a}\label{kappa}
\eeq
where $n_a$ is the density and $\mu_a$ the chemical potential of specie $a$.

As a second example we shall use
a Gau\ss{}-type potential
\beq
V_{G}(r)=V_0 {\rm e}^{-(r/\eta)^2}
\label{ggg}
\eeq
which has been used in nuclear physics applications\cite{D84,hsk95}
with $\eta=0.57 {\rm fm}^{-3}$ and $V_0=-453$ MeV.

We will proceed and derive analytical expressions which are
compared with the numerical solution of (\ref{vv}).
The numerical values are compared in table
\ref{tab1} with the solution of the KB equations. It is found an overall good agreement.

\subsection{Time dependent correlation energy}

We
calculate the  build up time of correlation, $\tau_c$
by inspecting the time derivative of the interaction energy.
We define $\tau_c$ as the time at which this
derivative becomes sufficiently small. This
corresponds to using (\ref{s2}) instead of
(\ref{s1}), but only with the first term according
to the finite
duration approximation of the last chapter. We have from (\ref{z1})
\beq\label{v2}
{\pa t} E_{\rm corr} &=& -\frac {1}{ 4 \hbar} \langle
\sin{{\Delta E
(t-t_0)\over \hbar}}
F(t) \rangle.
\eeq

\subsubsection{High temperature limit}
In the limit
of high temperature we neglect the
degeneracy and the equilibrium distribution takes the
Maxwell-Boltzmann form
\beq
f(p)=\frac n s \lambda^3 {\rm e}^{-{p^2 \over 2 m T}};\qquad
\lambda^2={2 \pi \hbar^2 \over m T}.\label{bol}
\eeq

We get for the Gau\ss{} potential and $b^2=\hbar^2/(2\mu T \eta^2)$ the result
\beq
&&{\pa t} E_{\rm corr}^G=\sum\limits_{ab}{4 n_a n_b \pi \eta^2 V_0^2 \over \sqrt{2 \mu T}
b^4} \left ({4 \mu \over M} \right )^2
\nonumber\\
&&\nonumber\\
&\times&{\pa \beta} \left ({\pi (\sqrt{\beta^2+16 t^2 T^2/\hbar^2}-\beta)\over 8 (\beta^2+16 t^2 T^2/\hbar^2)} \right )^{1/2}_{\beta
=2/b^2+4 t^2 T^2 /\hbar^2}\nonumber\\
&=& -\sum\limits_{ab}{3 n_a n_b \pi^{3/2} \eta^6 V_0^2 (2 \mu)^{3/2} \over 16 T^{5/2}} \left ( {4 \mu \over M} \right )^2
{1 \over t^4} + o(t^{-5}).\nonumber\\
&&\label{highe}
\eeq
We see
a monotonic decrease of the time derivative or
equivalently a monotonic increase of the correlation energy to
its equilibrium value. It is remarkable that the long time limit is
entirely determined by the classical value. Obviously
no quantum effects enter the formation of correlation in the high temperature limit.

We can also calculate the time dependent formation for Yukawa like potentials.
The time derivative of the correlation
energy leads to [appendix (\ref{i4})]
\beq
{\pa t} {E_{\rm corr}^Y(t) \over n}&=& -{e^2 \kappa T\over 2 \hbar}{\rm Im}
\left [(1+2 z^2 ) {\rm e}^{z^2} (1- {\rm erf} (z)) -{2 z \over \sqrt{\pi}} \right ] \nonumber\\&&
\label{class1}
\eeq
where we used $z =\omega_p \sqrt{t^2 - i t {\hbar \over T}}$ and
the collective (plasma)
frequency $\omega_p^2=\kappa^2 T/m$, compare with (\ref{kappa}).
This is the analytical quantum result of the time derivative of the formation of correlation. For the classical limit we
are able to integrate expression (\ref{class1}) with respect to times and arrive at \cite{MSL97a}
\beq
&&E_{\rm corr}^Y(t) -E_{\rm corr}^Y(0)= -{n \kappa e^2\over 4 } \nonumber\\
&\times&
\left [1+ {2 \omega_p t \over \sqrt{\pi}}-(1+2 (\omega_p t)^2 ) {\rm e}^{(\omega_p t)^2} (1- {\rm erf} (\omega_p t)) \right
].
\label{class2}
\eeq

It shows that the formation of correlations is basically given by the inverse
collective frequency $\omega_p$. Therefore during the first
stage of relaxation the fluctuating and collective effects are more important
than collisions.

\subsubsection{Low temperature limit}

The low temperature value is of special interest,
because it leads to a natural definition of the build up of
correlations.
Using the same steps as in appendix (\ref{low}) one
obtains from (\ref{v2})
\beq\label{v3}
{\pa t} E_{\rm corr} &=&
-\frac 1 2 m^4 p_f \langle {V^2 \over \cos{\frac \theta
2}} \rangle \tilde I_f
\label{to1}
\eeq
with abbreviation as in (\ref{vcos}).
It is now easy to perform the time integral in (\ref{to1})
to obtain the time dependence of the correlation
energy
\beq
&&E^{low}_{\rm corr}(t)-E_{\rm corr}^{low}(0)= E_{\rm corr}^{\rm low}
(1 +  \frac 1 3 ({ \epsilon_f+\epsilon_c \over \pi T})^2)^{-1}
\nonumber\\
&\times&
\left \{ 1-\frac 1 x \sin(x)
+\left ({\epsilon_f +\epsilon_c \over \pi T}\right )^2 \left (\frac 1 3 + \left [\frac 1 x \sin(x)\right ]''\right )\right \}
\label{corrt}
\eeq
with $x=2{\epsilon_f+\epsilon_c  \over \hbar} t$ and the equilibrium correlation energy
$E_{\rm corr}^{\rm low}$ from
(\ref{equil}) respectively.
This shows that the correlation energy
is built up and oscillates around the equilibrium value.
This oscillations are damped
with $t^{-1}$ in time.

We would like to point out here that the result for the
formation of correlations at low temperatures (\ref{corrt}) is
independent of the interaction used. Because higher
order interaction described by higher order diagrams
can be cast into a Boltzmann- like collision integral
with more involved transition matrix elements we have always the
same time dependence of
(\ref{corrt}) for binary interactions however with different
equilibrium correlation energy $E_{\rm corr}$.

\subsection{Numerical results}

\vspace{2ex}
\begin{figure}
\epsfxsize=8cm
\epsffile{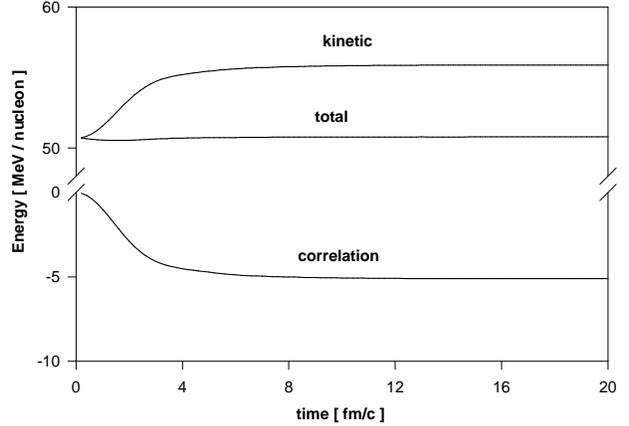}
\caption{\label{1}The time dependent kinetic and
correlation energy vs. time for a
counter-flowing streams of nuclear matter from
(\protect\ref{short1}). The
temperatures and densities of the
colliding beams are $T_1=10$ MeV, $n_1=n_o/60 $ and
$T_2=5$ MeV
$n_2=n_o/10$ and the relative momentum $1.5 \hbar/fm$,
which
corresponds a colliding energy of $45$ MeV/nucleon. The beams
are starting to interact at time point
$t_0$. One sees the build up of correlation energy
during a correlation-time  $\tau_c=3 fm/c$. The parameter are chosen in
such a way that the system is degenerate, which is
described by $n \lambda^3=0.416 <1$.
The total initial kinetic
energy corresponds (neglecting correlation energy)
to an equilibrated system with a temperature
of $T=32$ MeV.}
\end{figure}

Figure \ref{1} shows the time development of
the
kinetic energy as well as the correlation energy. The equation
(\ref{short1}) has been solved numerically
for two initially
counter-flowing streams of nuclear matter, where the
nucleons interact via the Gau\ss{} type of potential
(\ref{ggg}). The initial temperatures and densities of the
colliding beams are $T_1=10$ MeV $n_1=n_o/60 $ and
$T_2=5$ MeV
$n_2=n_o/10$ respectively. The relative momentum is $1.5 \hbar/fm$,
which
corresponds to a colliding energy of $45$ MeV/n.
We observe a
build up of correlations during the initial $3-4
fm/c$. Total energy is conserved and the
kinetic energy is increased by the same amount as the
correlation energy is decreased. Similar results has been found for a different system in \cite{BKSBKK96}.
This is because the system is initially prepared
to be uncorrelated at $t_0=0$.
If
time $t_0$ i.e. the time when the system is uncorrelated is shifted to
$t_0=-\infty$ i.e.  infinite
past, we would not observe any build up of
correlations. The equation
(\protect\ref{short1}) would then in fact reduce to the Boltzmann
equation.

\vspace{8ex}
\begin{figure}
\epsfxsize=8cm
\epsffile{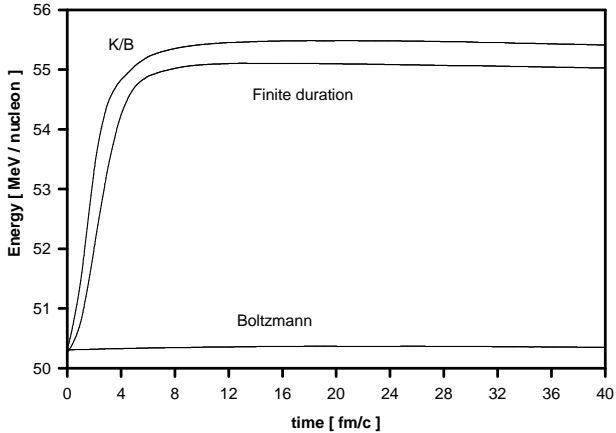}
\caption{\label{3}The time dependent kinetic energy from a
solution of the Kadanoff-Baym equation (KB) is shown
together with the results from the finite duration
approximation
(\protect\ref{short1}) and the Boltzmann equation.
For Boltzmann transport the kinetic energy is conserved in each
collision and therefore globally.
The conclusion is that the broadening of the
$\delta$
function of energy conservation in the finite duration approximation
almost accounts for the
time dependent built up of correlations from the exact KB
equation. }
\end{figure}

In figure \ref{3} we compare the results with the exact
solution of Kadanoff-Baym equations \cite{hsk95,hsk96}. We see
that the
finite duration approximation reproduces the exact
result
quite nicely. The small deviation is due to higher
order effects.

\begin{figure}
\epsfxsize=8cm
\epsffile{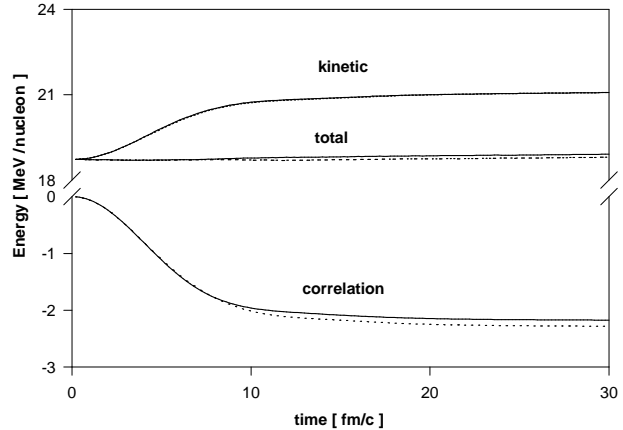}
\caption{\label{5}The time evolution of the correlation
and kinetic energy for two
initially counter -flowing streams of
nuclear matter with density and temperature $n_1=n_o/60
\quad
T_1=0.5 $ MeV
and $n_2=n_o/20 \quad T_2=0.1$ MeV moving with relative
velocity of $1
\hbar/fm$ which corresponds to a collision energy of
$21$ MeV/n. In this case the correlation time $\tau_c=9 fm/c$ which is
appreciably
larger than in Fig \protect\ref{1}. We ascribe this
to the smaller
thermal velocity of the particles in the present case.
The parameters are here in contrast to
figure \protect\ref{1} chosen such that the system is
degenerate $n \lambda^3=1.14 \ge 1$.
The equilibrated temperature is here
(neglecting correlations)  $T=11.6$ MeV.}
\end{figure}

In order to investigate a situation with higher degeneracy
we choose a model of two
initially counter -
flowing streams of
nuclear matter with density and temperature $n_1=n_o/60
\quad
T_1=0.5 $ MeV
and $n_2=n_o/20 \quad T_2=0.1$ MeV moving with relative
velocity of $1
\hbar/fm$ corresponding to a collision energy of
$21$ MeV/n. The interaction is again a Gau\ss{}-type of
potential.
  In figure \ref{5} is plotted the time
evolution of
the kinetic energy and the correlation.

\vspace{2ex}

This build up of correlations is independent of the
initial distribution form. If for example we choose a (equilibrium)
Fermi
distribution as the initial distribution
a build up of correlations will occur as well.
This is due to the fact that the spatial correlations relate in momentum
space to excitations,
resulting in a distribution looking somewhat like a Fermi-distribution
but with a temperature higher than that of the initial uncorrelated
Fermi-distribution.\cite{hsk95}

\begin{figure}
\parbox[t]{8cm}{
  \psfig{figure=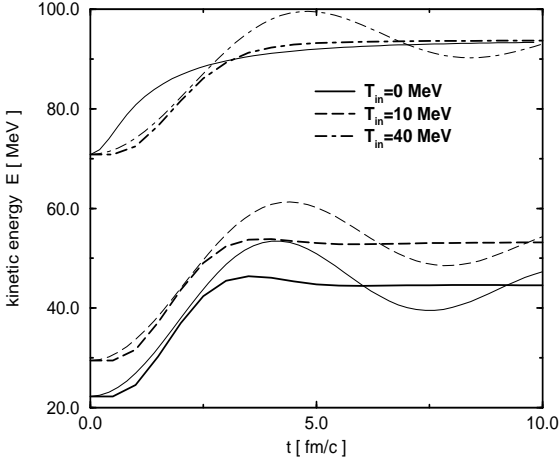,width=8cm,height=7cm,angle=-90}}
\caption{\label{ill}
The formation of correlation plotted as an increase of the kinetic energy with time for temperatures $1,10,40$ MeV. The
chemical potential is fixed to $37.1$ MeV which corresponds to
densities $0.16, 0.18,0.35 {\rm fm}^{-3}$ . The thick lines show results
from KB calculations while the thin lines
are approximate values via formula (\protect\ref{corrt}).
The equilibrium correlation energy was chosen to be equal to the
KB result. The oscillations are overestimated by the approximate formula. For $T=40$ MeV we plotted also the high temperature approximate value via (\protect\ref{highe}) as thin solid line. The built up of correlations is too fast.}
\end{figure}

In order to illustrate the temperature dependence of $\tau_{c}$
as well as to demonstrate the quality of limiting analytical
formulae, we plot in figure \ref{ill} (thick lines)
results from the solution of the Kadanoff
and Baym equations for a fixed chemical potential of $37.1$ MeV
and for three different temperatures. The figure shows
the increase of the kinetic energy (equivalent to the decrease of correlation
energy) with time.
The KB results are compared with those from
approximation (\protect\ref{corrt}). One sees that initially while
correlations are built up the agreement is good.
Especially at low temperatures the
oscillations discussed above are obvious in the approximate results
while the KB calculations only show a slight overshoot at the lowest
temperatures.
We believe that the discrepancy is due to the
neglected damping (and perhaps due to the necessary approximations used
in the integrations as discussed above).
The opposite approximate formula for high temperatures (\ref{highe}) is plotted also for the $T=40$ MeV case. The built up of correlations is too fast according this formula.
\vspace{2ex}

From the numerical inspection of this section we
summarize three facts:
(i) the build up of correlations is monotonic and
reaches the final value smoothly, (ii) the finite
duration approximation (\ref{short1}) is an excellent
approximation and (iii) also for initial Fermi distributions we see
the same build up of correlations. The latter point shows that this build
up of correlation is not due to an equilibration of an initial nonequilibrium
distribution but mainly due to the decay of higher order
correlation function which are condensed in the off-shell $sin$ -
factor in the collision integral (\ref{kinetic}).

\subsection{Correlation time}

From the numerical observation we conclude that the
correlation energy (\ref{vv}) is increasing monotonously with time
until it reaches its almost final value (\ref{eq}).
We assume
that the main formation time of correlations
$\tau=t_c-t_0$ is given by reaching this asymptotic limit.
This time can be estimated by the condition
\beq
E_{\rm corr}(\tau) &=& \frac 1 4 \langle
\frac{\cos{{\Delta E
\tau\over \hbar}}-1}{\Delta E}
F(t_c) \rangle \nonumber\\
&\approx& \frac 1 4 \langle \frac{1}{\Delta E}
F(\infty) \rangle
\equiv  E_{\rm corr}.
\label{conditio}
\eeq
We solve this equation for $\tau$ approximately
by replacing the $\cos$-function by a linear
approximation within the build up time interval
$t-t_0=(0,\pi \hbar /\Delta E)$
\beq
{1-\cos{{\Delta E (t-t_0) \over \hbar}}\over \Delta E}
\approx {2 (t-t_0) \over \pi \hbar} {\rm Sgn}(\Delta E).
\eeq
This approximation is correct in three points, the
initial time, the final formation time where
$<V(\tau)>$ has its maximum and a time point half of
this maximal time at $\pi \hbar /2/\Delta E$. The
latter point coincides too because $1-\cos{x t}$ has
there a turning point. Therefore this linear
approximation overestimates the function in the first
half of the interval $t-t_0=(0,\pi \hbar /\Delta E)$ and
underestimates it in the second part. Furthermore we use
equilibrium distribution functions. With the help of
this approximation we can solve (\ref{conditio})
\beq
\tau &\approx& \frac 1 2 \pi \hbar {\langle \frac{1}{\Delta E}
F(\infty) \rangle \over \langle
{\rm Sgn}(\Delta E) F(\infty) \rangle}.
\label{tau}
\eeq
If we use $<{\rm Sgn} (\Delta E) F>\approx E_{\rm corr} <\Delta E>$, where $<\Delta E>$ is
the mean transition
energy of the collision, we obtain $\tau \approx {\hbar
\over <\Delta E>}$.
This gives the intuitive picture of an {\it uncertainty} principle, i.e. a
smallest time scale determined by $\hbar$ divided by the transition
energy.

The high temperature value of (\ref{tau}) can be
calculated analogously to appendix \ref{high}. Instead of
(\ref{v}) and (\ref{vg}) we have now
\beq
{\pi \hbar \over 2 \tau}=4 T {\int\limits_0^{\infty} dx x^2
V^2(x)\int\limits_{-\infty}^{\infty}ds \,{\rm Sgn}(x (s+x)) {\rm
e}^{-s^2}
\over
\int\limits_0^{\infty} dx x
V^2(x)\int\limits_{-\infty}^{\infty}ds {{\rm
e}^{-s^2}\over x+s}},
\label{max}
\eeq
with $V(x)=1/(x^2+b^2)$ for Yukawa potential and $V(x)=\exp{-(x/b)^2}$
for Gau\ss{} potential.
For the latter we obtain with the help of appendix \ref{b}
[(\ref{bb})]
\beq
\tau&=& {\pi^2 \hbar \over 4 T} {1\over (2+b^2) (\frac \pi 2+
{\sqrt{2} b \over 2+b^2}- {\rm arctan}(\sqrt{2}/b))}\nonumber\\
&=&{\pi^2 \eta \over 8 v_{th}}(1- { b^2 \over 6}+ o(\hbar^4))
\label{max1}
\eeq
with the thermal velocity $v_{th}^2=T/\mu$ and
$b=\hbar/\eta/\sqrt{2 \mu T}$.

We see that in the low density or quasi-classical limit
$b\rightarrow 0$ the formation time of correlations are
determined entirely by the range of potential $\eta$ divided by
the thermal velocity.
This result is intuitively appealing.

The opposite limit of low temperatures can also be
evaluated like in appendix \ref{low}. This limit is
independent of the used interaction because following
(\ref{to}) the interaction part cancels out in
(\ref{tau}). Then we end up with Fermi integrals
which are similar to (\ref{c1}) and the resulting formation time of correlations is via (\ref{tau})
\beq
\tau_{\rm low T}&=&{\pi \hbar \over 3 \epsilon_f+\epsilon_c}
\left ( 1 + ({\pi T \over \epsilon_f+\epsilon_c})^2 \right) .
\label{landau}
\eeq
This time agrees with the time where the correlation
energy (\ref{corrt})
has reached its first maximum
\beq
\tau_c\approx {2 \hbar \over
\epsilon+\epsilon_c}.
\label{exact}
\eeq

This correlation time limits the validity of quasiparticle picture which is established at times greater than this \cite{MSL97a}.
Incidentally, in the early 1950s the criterion $
\hbar/k_BT<\tau$
was supposed to limit the validity of the Landau
Fermi-liquid
theory for metals \cite{P55}. Later it was
shown by Landau that
this criterion is irrelevant and he proposed the
correct
criterion $\tau > \hbar/\epsilon_F$. Here we have
explicitly calculated the formation time of
correlations which is found to be equivalent to the
memory time.
It is to remark that this result describes
just the
memory- or
collision duration time $\tau_{mem}\equiv1/E$, see
\cite{MR95}. For nuclear matter at saturation density this time is about $4-5$~fm/c and agrees with the numerical result of memory time \cite{GWR94}.

\subsection{Range of validity}

\begin{figure}
\epsfxsize=8cm
\epsffile{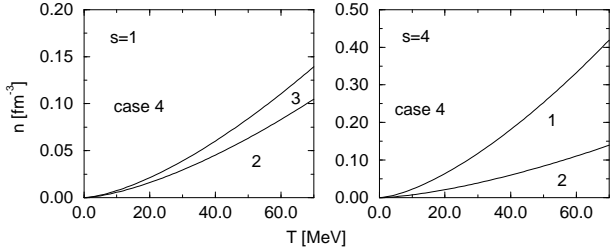}
\caption{\label{range}
The 4 different areas of density- temperature range according to (\ref{mat}). The high
temperature expansion is applicable in case 2 and the low temperature
expansion in case 4. The right figure represents the situation for
spin-isospin degeneracy $s=4$ and the left figure the situation for $s=1$ for comparison.}
\end{figure}

The validity of the low and high temperature expressions
can be discussed with the help of two parameters, the
value of the degeneracy $n\lambda^3=\exp{\mu/T}$ and the ratio
$\epsilon_f/T$. There are 4 cases

\beq
\matrix{case \;1 & n\lambda^3>1\qquad T>\epsilon_f & only\;for\;s>1\cr
case \;2 & n\lambda^3<1\qquad T>\epsilon_f & \cr
case \;3 & n\lambda^3<1\qquad T<\epsilon_f & only\;for\;s=1\cr
case \;4 & n\lambda^3>1\qquad T<\epsilon_f & \cr}
\label{mat}
\eeq
where $s$ is the degeneracy. In figure (\ref{range}) we plot this case for
nuclear matter with $m=938$ MeV.

\vspace{2ex}
It is clear that the low temperature expansion can only hold for case 4
and the high temperature expansion in case 2. For nuclear matter with
$s>1$ we see that the case 1 is not covered at all by these expansions.

\subsection{Discussion}
We compare in Table \ref{tab1} the different expansions with each other
as well as with exact results from KB-calculations.
The general agreement of the numerical value of the correlation energy (\ref{eq}) with the KB result is
striking. The correlation time (\ref{tau}) calculated numerically agrees as well reasonably.

For the case 2 in row 1 (for a high temperature and nondegenerate system)
we can
reproduce the build up time quite well with the approximate
formula (\ref{max1}) . This
becomes worse if we approach the degenerated condition $n \lambda^3=1$ in
second line of the table. Both of the high temperature cases are too narrow to
this condition to reproduce the correlation energy correctly. The high
temperature approximate formula (\ref{vtg}) overestimates the exact
value
 $ E_{\rm corr}$
obtained from the solution of the
time dependent kinetic equation. However in the case 4 for
low temperature and degenerated systems we can reproduce the simulation
value quite well. The build up time is slightly overestimated by the exact
low temperature value (\ref{exact}). The case 1 is
not covered by the approximate values which is demonstrated in the 5.
row of the table. There the correlation energy is much overestimated which is
quite clear from the discussion of Fig \ref{range}.
The systematic slight increase of the correlation time with temperature (for low temperature limit) is explained by the low
temperature expansion.

The results related to figure \ref{ill} are represented in rows
6-8. It underlines the general good description of case 4
and the failure of case 1.
The chemical potential was kept constant initially in these three
cases and the temperature
gradually increases, such that the chemical potential decreases.
According to ({\ref{exact}}) the correlation time should increase if
using the final chemical potential instead of the initial one.
This is clearly not the case in K/B solutions, where the
correlation time stays almost constant.
This shows clearly that the formation time cannot be completely
described by equilibrium formulae as done above. The correct formation
time is described by a chemical potential somewhat between initial and
final value.

We would like to repeat here that the expression
(\protect\ref{corrt})
for the low temperature limit is universal for
binary
collision approximations, independent of the interaction. The high temperature limit should also be
correct because the Born approximation used here is believed to be a good
approximation for fast particles. The intermediate region is quite open.
We have calculated only in Born approximation. Here especially
higher order correlations beyond second Born should be
employed. In summary: For low temperatures approximate analytical
expressions can be given and
for higher temperatures the analytical second
Born approximation should be approached, while the
intermediate region is left for numerical investigations.

\section{Summary}

The gradient approximation of the kinetic equation in second order Born
approximation is investigated.
The interaction energy is derived within different expansions.
The equilibrium value of the correlation energy is obtained from the first
order gradient expansion.
This equilibrium value is calculated for Yukawa and Gau\ss{} type of
potentials and
the results are analytically given in high and low temperature limit.
For contact potentials we rederive the known result for the ground state correlation energy.

A finite duration approximation of the non Markovian collision integral
is proposed which follows from time dependent Fermi's
Golden Rule and which is in good agreement with the numerical solution of
complete
collision integral. Furthermore it leads to the correct equilibrium value.
Numerical comparisons are made with the solution of the complete
Kadanoff and Baym equation in Born approximation and with this finite
duration approximation.

The build up time of correlations is investigated and it is found that
the low temperature value is universal for any approximation at the
binary collision level. It is shown that the formation time of
correlations is nearly determined by a ratio of $\hbar$ to the transfer
energy which can be considered as an analogue to uncertainty principle.

The high temperature limit shows roughly the time scale
that particle needs to travel through the potential range.
The validity of both the high and the low
temperature estimates are confirmed by numerical comparisons with
KB-results.

The time scale we are describing are just the life time of fireball assumed 
in the early stage of nuclear collision. This means the extracted temperatures 
from final stage products are usually wrongly extrapolated to early stages 
of nuclear reactions. This should effect the conclusions towards the caloric curve which is much discussed recently. Also the size and lifetime of hot reaction centers, which are extracted by interferometry methods should be critically revised as demonstrated in \cite{MK98}.

The authors like to thank P. Lipavsk{\'y} and V. {\v S}pi{\v c}ka for interesting discussions and A. Sedrakian for helpful comments.

\appendix

\section{Equilibrium correlation energy}\label{appc}

From the gradient expansion (\ref{s1}) we see that an
equilibrium value of the correlation energy
$E_{\rm corr}^{\rm eq}=<V_0 F_0>$ is approached for large
times. Here $F_0$ is given by Eq. (\ref{short2}) with equilibrium
distribution functions and $V_0$ is given by
(\ref{s11}) such that we have
\beq
&&E_{\rm corr}^{\rm eq}(\infty)-E_{\rm corr}^{\rm eq}(0)=
-<\frac 1 4 \frac {P}{ \Delta E
}F_0>\nonumber\\
&=&
-\frac{s_1 s_2}{\hbar }
     \int \frac{dp_1 dp_2 dp_1' dp_2'}{(2 \pi \hbar
)^9}
     V(\mid p_1 - p_1'\mid )^2
          \nonumber \\
  &\times& \delta (p_1 + p_2 - p_1' - p_2')
f_0(p_1') f_0(p_2')
{\bar f_0}(p_1)
{\bar f_0}(p_2)
     \nonumber \\
  &\times & {{\rm P} \over E_1 + E_2 - E_1' - E_2'}
\label{eq}
\eeq
where in the second equality we have employed symmetry
relations resulting in a factor 2 and have written
all terms explicitly.

We are now able to calculate the correlation energy
analytically. In order to compare with known results we calculate first the ground state energy for contact potentials.
Then we will be able to find analytical expressions for two extreme cases, i.e. the high
temperature and low temperature limit.

For performing the integrations in the collision integral
we define new coordinates by
$\vec p= \vec
p'_1-\vec p_1$, $\vec \Pi={m_b \over (m_a+m_b)}(\vec
p_1+ \vec p'_1)$ and $\vec p'= \vec p_2-\vec p'_2$, $\vec
\Pi'={m_b \over (m_a+m_b)}(\vec p_2+ \vec p'_2)$.
For the
$\Pi,\Pi'$ coordinates we
use cylindrical coordinates $\vec \Pi=\vec
\Pi_{\rho}+\vec \rho$ where $\vec\Pi_{\rho}$ is
parallel to the z-axis fixed by $\vec p$.  Within
these coordinates the momentum conservation takes the
simple form $\vec p'=\vec p$ and the energy difference is
$\Delta E={p\over 2 \mu} (\Pi_\rho'-\Pi_\rho)$ with
$\mu$ the reduced mass.

For the contact potential we find from (\ref{eq}) the known ground state correlation energy \cite{PN66}
\beq
{E_{\rm corr}\over N}=\frac 3 5 \epsilon_f {4 (2 \log 2-11)\over21 \pi^2} ({p_f a\over \hbar})^2
\label{contact}
\eeq
with the scattering length $a$ related to the coupling constant of the contact potential $4\pi \hbar^2 a/m$.
As pointed out in \cite{LL79} we have to subtract an infinite value, i.e. the term proportional to $f_1 f_2$ in
(\ref{eq}). This can be understood as renormalization of the contact potential and is formally hidden in
$E_{\rm corr}^{equil}(0)$ on the left hand side. For finite range potentials we have an intrinsic cut-off due
to range of interaction and such problems do not occur. For the standard procedure at low temperatures when
the angular and energy integrals are separated, we have to care about this cut-off once more because during
this procedure we neglect the effect of potentials but we will have to integrate off-shell.

\subsection{High temperature limit \label{high}}
The high temperature limit is performed using the
Maxwell-Boltzmann distribution (\ref{bol}) for the calculation of
the correlation energy (\ref{eq}).

\subsubsection{Yukawa-type potential}

Performing a series of trivial integrals we obtain the
expression for the correlation density for the Yukawa potential
\beq
E_{\rm corr}^{\rm Y}&=& -\pi \sum\limits_{a b}({4 \mu \over m_a+m_b})^2 {
g_{ab}^2 n_a n_b \over \kappa T} I_1[b] \nonumber\\
I_1[b]&=&{2 b \over \pi^{3/2}} \int\limits_0^{\infty}
{x dx \over (x^2+b^2)^2}
\int\limits_{-\infty}^{\infty}{d s \over s+x}{\rm
e}^{-s^2}
\label{v}
\eeq
where the abbreviation $b^2=(\hbar \kappa)^2/8 \mu T$
has been used. The parameter $b$ proportional to
$\hbar$ is responsible for quantum corrections.
The last integral in (\ref{v}) is performed in appendix
\ref{b} and the final result reads
\beq
E_{\rm corr}^{\rm Y}&=& -\pi \sum\limits_{ab}({4 \mu \over m_a+m_b})^2 {
g_{ab}^2 n_a n_b \over \kappa T} \left (1-\sqrt{\pi} b \,{\rm
e}^{b^2} {\rm erfc} (b)\right ).\nonumber\\
\label{vt}
\eeq
This result is the correlation energy in second Born
approximation. The quantum corrections
are condensed in the $b$ dependent function.  For identical plasma
particles ($g=e^2/\epsilon_0$) the result reads
\beq
{E_{\rm corr}^Y \over n}&=&{e^2 \kappa \over 4
\epsilon_0} \left (1-\sqrt{\pi} b \,{\rm e}^{b^2}\,{\rm erfc} (b)\right )\nonumber\\
&=&{e^2 \kappa \over 4
\epsilon_0} (1-\sqrt{\pi} b+2 b^2 + o(\hbar^3)).
\eeq
The
classical limit shows half of the exact low density limit of
correlation or self energy, \cite{kker86}, page 115, which we calculate in (\ref{i11}) in appendix \ref{b}
\beq
E_{\rm corr}^{Debye}&=&{e^2 \kappa \over 2
\epsilon_0}{\sqrt{\pi} \over b} \left (1-{\rm e}^{b^2} {\rm erfc}
(b)\right )\nonumber\\
&=&{e^2 \kappa \over 2
\epsilon_0}(1-{\sqrt{\pi} \over 2} b+\frac 2 3 b^2 + o(\hbar^3)).\label{class}
\eeq

This difference follows from the static approximation of the screened
interaction by the Debye potential, $W(\omega)\approx W(0)\equiv
V_{\rm D}$. The energy dependency of true $W$ reflects that the
screening of Coulomb potentials is a dynamical process. The Debye
potential is a correct approximation for the scattering rates, since
the transferred energy $\omega$ is small being limited by the on-shell
condition $\Delta_E=0$. For the off-shell
processes, the transfered energy can reach arbitrary high values for which
the Debye potential would fail. Instead we would have to use the dynamical screened potential
$W(q,{q (q-2 k) \over 2 m_a})$
in the Levinson equation (\ref{kinetic}). At low energies, the imaginary part of $W$
needed in the scattering integral reads ${\rm Im}W=|W|^2{\rm Im}\Pi
\approx V^2_{\rm D}{\rm Im \Pi}$. The principle value integration in the energy function (\ref{eq})
can be evaluated via Kramers-Kronig relations between the real and the imaginary part of
the screened potential. The conjugated real part is
${\rm Re}(W-V_{\rm C})=V_{\rm C}{\rm Re}\Pi W\approx V_{\rm C}
{\rm Re}\Pi{\rm Re}W\approx V_{\rm C}{\rm Re}\Pi V_{\rm D}$, where
$\Pi$ is the susceptibility.  Thus the dynamical screened potential would lead to $V_{\rm C}
V_{\rm D}\sim{\rm e}^{-\kappa r}r^{-2}$, while the static approximation
yields $V^2_{\rm D}\sim{\rm e}^{-2\kappa r}r^{-2}$. The factor of two in
the exponent brings the factor of two into denominator in formula
(\ref{vt}).
The corresponding integrals leading to (\ref{class}) can be found in appendix \ref{b}. This means that we do not consider the
formation of effective Debye potential, but use it here as an effective one present from the beginning.

\subsubsection{Gau\ss{}-type potential}

For the Gau\ss{} potential the same steps as before lead instead of (\ref{v})
to the expression
\beq
E_{\rm corr}^{G}&=& -{({4 \mu \over m_a+m_b})^2  \pi^{3/2}
n_a n_b \eta^3 V_0^2 \over 2 T \sqrt{2}} I_2[b],
\nonumber\\
I_2[b]&=&{2^{3/2} \over \pi b^3} \int\limits_0^{\infty}
x {\rm e}^{-2(x/b)^2} dx
\int\limits_{-\infty}^{\infty}{d s \over s+x}{\rm
e}^{-s^2},
\label{vg}
\eeq
where $b^2=\hbar^2/(2 \mu T \eta^2)$. With the help of
(\ref{bb}) in appendix \ref{b} we obtain
\beq
E_{\rm corr}^{G}&=& -({4 \mu \over m_a+m_b})^2 {\pi^{3/2}
V_0^2 \eta^3 n_a n_b \over 2 \sqrt{2} T} {1 \over 2
+b^2}.
\label{vtg}
\eeq

\subsection{Low temperature limit \label{low}}
In the opposite limit of small temperatures compared to
the Fermi energy we employ the well known methods of
Fermi liquid theory \cite{SHJ89,BP91}. For simplicity
we now use only one type of particles. The angular and
energy integrals in (\ref{eq}) can then be separated with
the result
\beq
E_{\rm corr}=-\frac 1 2 m^4 p_f \langle {V^2 \over \cos{\frac \theta
2}} \rangle I_f,
\label{to}
\eeq
where
\beq
&&\langle {V^2 \over \cos{\frac \theta 2}}\rangle ={1
\over (2 \pi \hbar)^9} \int {d \Omega_1 d \Omega d
\phi_2 \over \cos{\frac \theta 2}} V^2( p_1-p_1')
\label{vcos}
\eeq
with standard notation of angular
integrals \cite{SHJ89}.
The Fermi integral $I_f$ is done in appendix \ref{c}
and the final result for the Yukawa- or Gau\ss{}-
potential reads
\beq
E_{\rm corr}&=&\frac 1 6 s_1 s_2 \tilde \epsilon_f \left (T^2+ \frac 1 3 \left ({4 \tilde \epsilon_f \over \pi}\right
)^2\right ) \left ({m \over \hbar^2}\right )^4
\nonumber\\
&\times& \left \{\matrix{ {g^2\over 2 \kappa^3 \pi^2} \left (
\arctan{\frac {1}{ b_l}}+{b_l \over 1+b_l^2} \right )\quad
Yuk.\cr
{V_0^2 \eta^5 \over 2 \sqrt{2 \pi}} {\rm erf}({p_f \eta \sqrt{2} \over \hbar})\quad \mbox{\it
Gau\ss{}}} \right .\nonumber\\
&&
\label{equil}
\eeq
where $\tilde \epsilon_f={\epsilon_f +\epsilon_c \over 4}$
and $b_l={\hbar \kappa \over 2 p_f}$.
The best choice for
cut-off we found $\epsilon_c \approx \epsilon_f$.
This result shows that even for zero temperature we obtain a
correlation energy due to ground state correlations
by Pauli blocking, which we have explicitly demonstrated for contact potentials (\ref{contact}).

It should be noted however that the temperature $T$ refers
to the final equilibrated distribution. It is, in general, not a
Fermi-distribution but is modified because of the correlations.
This is for example illustrated by the results shown in figures 1 and 3
in \cite{hsk95}.

\subsection{Numerical values}

\begin{figure}
\epsfxsize=8cm
\epsffile{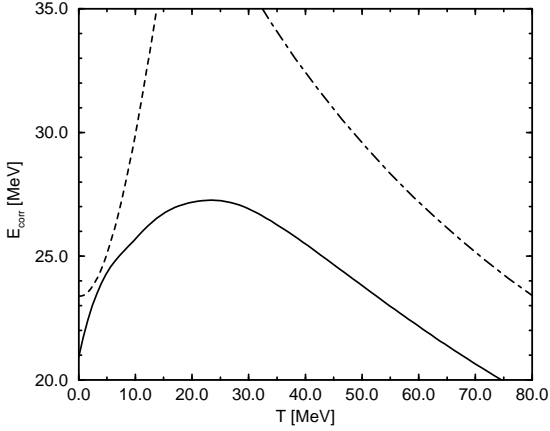}
\caption{\label{e}The correlation energy (\protect\ref{eq}), solid line, is compared with the low (\protect\ref{equil}),
dotted line,  and the high temperature (\protect\ref{vtg}), dot-dashed line,
approximations. Results are shown as a function of temperature at a
nuclear density of $n=0.18 {\rm fm}^{-3}$. See also discussion in text.}
\end{figure}

In order to check these low and high temperature approximations,
we plot in figure \ref{e} the  results from
(\ref{equil}) and (\ref{vtg}) with the numerical result of (\ref{eq})
for the Gau\ss{}-potential. We have chosen a
normal nuclear density, $n=0.18 \, {\rm fm}^{-3}$.
We see that the low temperature expansion,
(\ref{equil}), gives reasonable agreement with (\ref{eq}) at low temperatures
only.
The high temperature limit is approached actually
only at very high temperatures beyond 100
MeV at this density. We would like to emphasize that the numerical
computation of
of (\protect\ref{eq}) involves a complicated
four - dimensional principal value integration. While we consider the high
temperature limit to be accurate, the low temperature limit is approximate and dependent on the energy integration. The cut-off
value of the energy integration (see discussion after
(\protect\ref{eq}))
was chosen to be $\epsilon_f$ throughout the calculation.
In the low temperature limit we
used instead of $\epsilon_f$ the Sommerfeld expansion
$\mu(T)=\epsilon_f \left( 1-{\pi^2 \over 12} \left ({T \over \epsilon_f} \right )^2 \right ) +o(T^3)$.

\vspace{2ex}

\section{Weakening of initial correlation and the limit
of complete collisions}\label{append}

In order to derive kinetic equations for the long time
evolution one uses the limit
$t_0 \rightarrow -\infty$ where $t_0$ is the initial
time where correlations beyond the one-body are assumed to be absent.
Two-
particle correlations vanish then in the remote past.
>From (\ref{short}) it is obvious that we could consider
$t \rightarrow \infty$ alternatively. This is sometimes referred as \it complete collisions \rm, i.e.
uncorrelated asymptotic states, which can include one-body correlations,
of course .

We like to show now that special care is needed when
this limit is
carried out,
especially, if one likes to consider non-markovian
corrections as we did within
the
gradient approximation (\ref{s1}). Especially we will
find that the ordering of the gradient expansion and this long time
limit are not easily interchangeable.

To illustrate this fact let us first carry out the $t_0
\rightarrow -\infty$
limit
on the zero and first order terms $V_{n}(t)$ in Eq. (\ref{s1}).
This implies that we consider the limit of complete collisions.
The result is
\beq
V_0(t)&=&-{1- \cos{{ \Delta E (t-t_0)\over \hbar}}
\over 4 \Delta E} \rightarrow -\frac 1 4 \frac {\rm P}{
\Delta E}, \nonumber\\
V_1(t)&=&-\frac \hbar 4 {\pa \Delta E}{\sin{{ \Delta E
(t-t_0)\over \hbar}} \over \Delta E} \rightarrow -\frac
\hbar 4 \delta'(\Delta E),\nonumber\\
...\nonumber\\
V_n(t) &\propto& \hbar^n,
\label{s11}
\eeq
where ${\rm P}$ is the principal value as given in the
appendix.

Therefore we obtain in the long time limit (i.e. for completed
collisions)
\beq
{\pa t}\lim_{t_0 \to -\infty}E_{\rm corr}=-<\frac 1 4 \frac
{\rm P}{ \Delta E} {\pa t} F(t)> +o(F''(t))
\nonumber\\
\label{r1}
\eeq
On the other hand we can \it first \rm perform the time
derivative and \it then \rm the long time limit. To this end we
employ the gradient expansion of the collision integral
(\ref{s2}) and obtain with the help of (\ref{z1})
\beq
{\pa t} E_{\rm corr}&=& <I_0 F(t)> +<I_1 F'(t)> +
o(F''(t))\nonumber\\
&=&-\frac 1 4{\pa t} <\left [ {1- \cos{ {\Delta E
(t-t_0) \over \hbar}} \over \Delta E} F(t) \right ]>
\nonumber\\
&&+ <\frac {1}{ 4 \hbar} (t-t_0) \sin{{ \Delta E
(t-t_0)\over \hbar}} F'(t)>\nonumber\\
\rightarrow
-{\pa t}&& <\frac 1 4 \frac {\rm P}{ \Delta E}
F(t)>-<\frac 1 4 \epsilon {\pa \epsilon} \frac {\rm P}{
\Delta E} F'(t)>,\nonumber\\
\label{r11}
\eeq
where the relations
$T \cos{x T}\rightarrow
 -\epsilon \frac{\partial}{\partial \epsilon}\pi
 \delta^{\epsilon}(x)$ and $
T \sin{x T} \rightarrow -\epsilon \frac{\partial}{\partial
\epsilon}
\frac{{\rm P}}{x}$ has been used.
We see that we obtain with the help of $x \frac{{\rm P'}}{x}= -\frac{{\rm P}}{x}-\epsilon
\frac{\partial}{\partial
\epsilon}\frac{{\rm P}}{x}$
\beq
\lim_{t_0 \to -\infty} {\pa t} E_{\rm corr}=\frac 1 4
<\Delta E \frac {\rm P'} {\Delta E
}F'(t)>+o(F''(t)),\nonumber\\
\label{r2}
\eeq
where ${\rm P'}$ is the derivative of the principal
value as given in the
appendix.
Comparing with (\ref{r1}) we recognize that
\beq
(\lim_{t_0 \to -\infty} {\pa t}-{\pa t}\lim_{t_0 \to
-\infty})E_{\rm corr}= \cr
-<\frac 1 4 \epsilon {\pa \epsilon}
\frac {P }{\Delta E} F'(t)>.\nonumber\\
\label{rul}
\eeq
Here we can deduce the rule
from (\ref{rul}) for large times
\beq
E_{\rm corr}^{\rm grad-exact}(t)>=E_{\rm corr}^{\rm
grad-coll}(t)+<\frac 1 4 \epsilon {\pa \epsilon}
\frac {P }{\Delta E} F(t)>
\label{rule}
\eeq
that the exact correlation energy in gradient
approximation for large times $E_{\rm corr}^{\rm
grad-exact}(t)$ is given by the correlation
energy from gradient expansion of collision integral
$E_{\rm corr}^{\rm grad-coll}(t)$ via an infinitesimal
correction term.
For any function $F$ with existing derivatives the distribution
$\epsilon {\pa \epsilon}
\frac {P }{\Delta E}$ vanishes. Therefore we are allowed to
identify the correlation energy from the kinetic equation in
gradient approximation with the corresponding correlation energy
in second Born approximation.
Total energy is conserved in first order gradient expansion for
nondegenerate systems \cite{KBBS96}. For the degenerate systems we can
prove complete energy conservation also for degenerate and space nonlocal
systems \cite{LSM97}.

However we like to point out that the above result
is not valid if the transition probability contains poles, e.g. bound
states or pairing ones. Then we get a nonvanishing last term in
(\ref{rule}) expressing long living correlations such as bound states or
pairing. The same discussion is applicable for formal divergent expressions as it appears e.g. for contact
potentials. With the help of (\ref{rule}) we can renormalize the expression to obtain the ground state energy
as we will demonstrate later.

\subsection{Long time limit of the collision integral}

The message of the last paragraph does not mean that
the long time limit $t_0
\rightarrow -\infty$
of the collision integral is not
uniquely defined. To show this we consider the two orderings of
operations of the time
derivatives and the limit of remote past respectively.
For this purpose
we  write the gradient expansion up to first
order of the kinetic equation (\ref{short}) in two different ways
\begin{eqnarray}\label{gradient}
{\pa t}f_1&=&\frac 1 \hbar \frac{\sin{{(t-t_0) \Delta E
\over \hbar}}} {\Delta E} F(t)\nonumber\\
&&+ \frac{\partial}{\partial
\Delta E}
\frac{\cos{{\Delta E (t-t_0)\over \hbar}}-1}{\Delta E}
F'(t)+o(F''(t))\nonumber\\
&=& \frac 1 \hbar \left (\frac{\sin{{(t-t_0) \Delta E
\over \hbar}}}{\Delta E} +{(t-t_0) \over \hbar}
\cos{{\Delta E (t-t_0) \over \hbar}}\right )
F(t)\nonumber\\
&&+   \frac{\partial}{\partial t} \left [
\frac{\partial}
{\partial \Delta E} \frac{\cos{{\Delta E (t-t_0) \over
\hbar}}-1}{\Delta E}
F(t) \right ].
\end{eqnarray}
The first equality is written in terms of the derivatives $F^{n}$.
Now we take the limit $t_0 \rightarrow -\infty$ on this and obtain
\begin{eqnarray}\label{1a}
{\pa t} f_1&=&\frac 1 \hbar
\frac{\epsilon}{\epsilon^2+(\Delta E)^2} F(t)
-\frac{\partial}{\partial t} \left [
\frac{\partial}{\partial \Delta E}
\frac{\Delta E}{(\Delta E)^2 +\epsilon^2} \, F(t)
\right]\nonumber\\
&\equiv&\frac \pi \hbar \delta(\Delta E) \, F(t) -
\frac{\partial}{\partial t} \left [
\frac{{\rm P'}}{\Delta E} \, F(t)\right ].
\end{eqnarray}
Equation (\ref{1a}) is just the generalized kinetic
equation derived in
\cite{MR95,SL95} which leads to quantum Beth-Uhlenbeck
virial corrections in
nonequilibrium. It consists of the usual Boltzmann
collision integral and
a correction represented by the second term on the right hand
side of Eq.
(\ref{1a}).

We next invert the order of operations, i.e. the time derivation and the long time limit, and therefore use
the second equality of (\ref{gradient})
where the time
derivative is in front of the
collision
integral.
Applying the limit
$t_0 \rightarrow -\infty$ one obtains
\beq\label{2}
{\pa t} f_1&=&\frac 1 \hbar
(\frac{\epsilon}{\epsilon^2+(\Delta E)^2}-
\epsilon \frac{\partial}{\partial \epsilon}
\frac{\epsilon}{\epsilon^2+(\Delta E)^2})
F(t)\nonumber\\
&-&\frac{\partial}{\partial t} \left [
\frac{\partial}{\partial \Delta E}
\frac{\Delta E}{(\Delta E)^2 + \epsilon^2} F(t)
\right].
\eeq
Whereas the second term is just the same principal
value expression
as in (\ref{1a}), the first term is at first glance different,
but one finds
\begin{eqnarray}
\frac{\epsilon}{\epsilon^2+(\Delta E)^2}-
\epsilon \frac{\partial}{\partial \epsilon}
\frac{\epsilon}{\epsilon^2+(\Delta E)^2}&=&
\frac{\epsilon}{\epsilon^2+(\Delta E)^2}
\frac{2 \epsilon^2}{\epsilon^2+(\Delta E)^2}\nonumber\\
&=&\pi \delta (\Delta E)
\end{eqnarray}
so that the two results
(\ref{1a}) and (\ref{2}) are in fact equivalent.

This means that despite the care which is necessary for
obtaining
the mean value of the correlation energy (\ref{rule}), the
first order gradient
expansion for the generalized kinetic equation
has a unique long time limit independent of the ordering of the two
operations involved.

\section{Error function integrals}\label{b}
The following integral is needed in (\ref{vt}) for the calculation of the correlation energy
\beq
I_1[b]&=&{2 b \over \pi^{3/2}} \int\limits_0^{\infty}
{x dx \over (x^2+b^2)^2}
\int\limits_{-\infty}^{\infty}{d s \over s+x}{\rm
e}^{-s^2}\nonumber\\
&=&-{2 b \over \sqrt{\pi}} \int\limits_0^{\infty} {x dx
\over (x^2+b^2)^2}
{\rm Im} \, w(-x+i \epsilon)\label{b1},
\eeq
where the complex error function has been introduced
\cite{a84}
\beq
w(z+i\epsilon)={i \over \pi}
\int\limits_{-\infty}^{\infty}{{\rm e}^{-t^2} \over
z-t+i \epsilon}.
\label{b22}
\eeq
Performing a partial integration and using $w'(z)=-2 z
w(z) +2 i/\sqrt{\pi}$ we get
\beq
I_1[b]&=&1+{2 b \over \sqrt{\pi}}
\int\limits_0^{\infty} {x dx \over (x^2+b^2)}
{\rm Im} \, w(-x+i \epsilon).
\eeq
Reintroducing (\ref{b22}) the last integral is easily carried out by
interchanging the two integrals and we obtain finally
\beq
I_1[b]
&=&1-b \sqrt{\pi} {\rm e}^{b^2} {\rm erfc} (b).\label{i1}
\eeq

In order to verify (\ref{class}) we have instead of (\ref{b1})
\beq
&&I_{11}[b]={2 b \over \pi^{3/2}} \int\limits_0^{\infty}
{dx \over x (x^2+b^2)}
\int\limits_{-\infty}^{\infty}{d s \over s+x}{\rm
e}^{-s^2}\nonumber\\
&=&{1 \over b \pi^{3/2}} \left ( \int\limits_{-\infty}^{\infty}
{dx \over x}
\int\limits_{-\infty}^{\infty}{d s \over s+x}{\rm
e}^{-s^2}\right .\nonumber\\
&&\left . -\pi b \int\limits_{-\infty}^{\infty}{d s \over s^2+b^2}{\rm
e}^{-s^2} \right ),
\eeq
where a simple decomposition has been performed. For the first part we apply the Poincar\'e- Bertrand theorem \cite{DDW90}
which states
\beq
&&\int {dx \over x-u} \int  {d y \over y-x} f(x,y) =\nonumber\\
&&\int dy \int dx {f(x,y) \over (x-u)(y-x)} -\pi^2 f(u,u)
\eeq
and obtain
\beq
I_{11}={\sqrt{\pi} \over b} (1-{\rm e}^{b^2} {\rm erfc}(b)).\label{i11}
\eeq

Next we perform the integral required in (\ref{vg})
where we put $c=2/b^2$
\beq
{\sqrt{\pi} b^3 \over 2^{3/2}} I_2[b]&=&-{1\over
\sqrt{\pi}} \int\limits_0^{\infty} dx x {\rm e}^{-c
x^2}\int\limits_{-\infty}^{\infty}{d s \over s-x}{\rm
e}^{-s^2}\nonumber\\
&=&-{1 \over \sqrt{\pi}} \int\limits_0^{\infty} {ds
\over s} {\rm e}^{-{c \over 1+c} s^2}
(f(s)-f(-s))\nonumber\\
f(s)&=&\int\limits_0^{\infty} dx x {\rm e}^{-(1+c) (x+
{s \over 1+c})^2}.
\label{b2}
\eeq
Here we have derived the last form by dividing the
integral $s $ into two parts corresponding to the pole
at $s=\pm x$ and performing variable substitution.
The integral $f(s)$ is elementary and yield
\beq
f(s)&=&-{s \over (1+c)^{3/2} }{\sqrt{\pi} \over 2} {\rm
erfc} ({s \over \sqrt{1+c}}) + {1 \over 2 (1+c)} {\rm
e}^{-{s^2 \over (1+c)}}.\nonumber\\
&&
\eeq
Substituting this into (\ref{b2}) we obtain the result
\beq
I_2[b]&=&{2^{3/2} \over b^3 (1+c)^{3/2}\sqrt{\pi}} \int\limits_0^{\infty}
{\rm e}^{-{c \over 1+c} s^2} ds\nonumber\\
&=&{\sqrt{2} \over b^3 \sqrt{c}} {1 \over
1+c}\nonumber\\
&=&{1 \over b^2 +2}.
\label{bb}
\eeq

In (\ref{class1}) we need the integral
\beq
I_4[b]&=&{4 b \over \pi^{3/2}} \int\limits_0^{\infty}
{x^2 dx \over (x^2+b^2)^2}
\int\limits_{-\infty}^{\infty}d s{\rm
e}^{-s^2} \sin{(s+x) x t}\nonumber\\
&=&{2 b \over \pi} \int\limits_{-\infty}^{\infty} {x^2 \sin{t x^2} dx
\over (x^2+b^2)^2}
{\rm e}^{-t^2 x^2/4}\nonumber\\
&=&-{1 \over \pi} {\pa b}{\rm Im} \int\limits_{-\infty}^{\infty} {x^2 dx
\over (x^2+b^2)}
{\rm e}^{-(t^2 /4 + i t) x^2}\nonumber\\
&=&
{\rm Im}
\left [(1+2 z^2 ) {\rm e}^{z^2} (1- {\rm erf} (z)) -{2 z \over \sqrt{\pi}} \right ], \nonumber\\&&
\label{i4}
\eeq
where we used $z =b \sqrt{t^2/4 - i t}$.

Another error function integral is used in (\ref{max1})
\beq
I_3[a]=\sqrt{\pi} \int\limits_0^{\infty}dx x^2 {\rm e}^{-a x^2} {\rm
erf}(x).
\eeq
To this end we use a tabulated integral \cite{RG} (6.285.1)
\beq
I_{31}[a]&=&\int\limits_0^{\infty}dx {\rm e}^{-a x^2} {\rm
erf}(x) \nonumber\\
&=&{\frac \pi 2 - {\rm arctan}{\sqrt{a}}\over \sqrt{\pi a}}
\eeq
The desired integral is then
\beq
I_3[a]&=&-\sqrt{\pi}{\pa a} I_{31}[a]\nonumber\\
&=&\frac{1}{2 a^{3/2}} \left ( \frac \pi 2 - {\rm
arctan}(\sqrt{a})+{\sqrt{a} \over (1+a)} \right ).
\eeq

\section{Calculation of collision integral}\label{c}

In order to calculate the collision integral for
Fermi functions in (\ref{to}) special care is needed.
The problem differs from the
usual calculations of collision integrals in that we
do not have energy conservation but an off-shell
principal value. Therefore the limit $\lambda = \mu /T
\rightarrow \infty$ cannot be performed straightforwardly.
The reason is that the energies are restricted to
the neighborhood of the Fermi level, but we have to allow
off-shellness due to the denominator of the energy expression.
We have used the approximation to separate the collision integral into
angular and energy integrations. This leads to formally divergent results because the matrix element of
interaction is not included in the energy integrals and the latter
become infinite. In order to cure
this defect
we restrict the energies available from above. Then the expressions become analytical and we fix the upper limit of
energy integration by matching to the numerical result found from (\ref{eq}). The best fit was found to be two times
Fermi energy, i.e. $\lambda_c=\lambda$.
We shortly sketch the derivation with this upper limit.

Introducing dimensional variables $x=(\epsilon-\mu)/T$
and $\lambda=\mu/T$ we have
\beq
I_f&=&T^3\int\limits_{-\lambda}^{ \lambda_C}{dx_1dx_2dx_dx_
2' \over
({\rm e}^{x_1}+1)({\rm e}^{x_2}+1) ({\rm e}^{-x_1'}+1)
({\rm e}^{-x_2'}+1) }\nonumber\\
&&\times{1 \over (x_1+x_2-x_1'-x_2')}.
\eeq
Now we switch to difference and center of mass
coordinates via $p=(x_1+x_2)/2$ and $r=x_1-x_2$ and
analogously for $p',r'$. The integrals over $r$ and
$r'$ are then trivially carried out because their
integration range is $(- (\lambda+\lambda_c),\lambda+\lambda_c)\rightarrow (-\infty,\infty)$.
In new
coordinates $x=p-p'$ and $z=p+p'$ the result reads
\beq
I_f&=&-\frac 1 2 T^3\int\limits_{-(\lambda+\lambda_c)}^{\lambda+\lambda_c}{dx \over
x}{\rm e}^{-x} \int\limits_{-2 \lambda}^{2 \lambda_c} dz
{x^2-z^2 \over \cosh{x}-\cosh{z}}\nonumber\\
&=&{2 \pi^2 T^3 \over 3} \int\limits_{-2(
\lambda+\lambda_c)}^{2( \lambda+\lambda_c)}{dt \over {\rm e}^{t}-1} (1+({t \over 2 \pi})^2)\nonumber\\
&=&- {4 \pi^2 T^3(\lambda++\lambda_c) \over 3} (1+ \frac 1 3 ({\lambda+\lambda_c \over \pi })^2)
\label{c1}
\eeq
where we have employed the integral (\ref{d1}). We see
the explicit dependence of this off-shell integral on
$T^3 \lambda=T^2 \epsilon_f$.

We would like to remark that the limitation of upper available energies is equivalent to cut- off procedures,
which has been employed within memory collision integrals. This cut- off comes here from the approximate
calculation of the collision integral and is fixed to reproduce the numerical
results.

\section{A useful integral}\label{d}

\begin{figure}
\epsfxsize=7cm
\epsffile{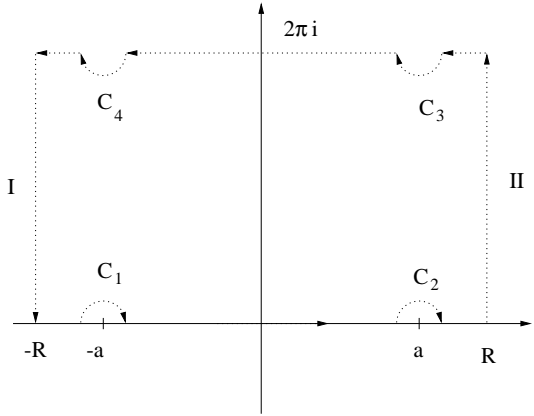}
\caption{\label{6}The integration contour for the
integral (\protect\ref{d1})}
\end{figure}

To evaluate the useful integral
\beq
J_n={\rm P} \int\limits_{-\infty}^{\infty} dx {x^n
\over \cosh{x}-\cosh{a}}
\eeq
we use the integration contour as depicted in figure
\ref{6}.
\vspace{2ex}

We have the following identity in the limit of large
$R$ and small radius around the poles
\beq
{\rm P} \int\limits_{-\infty}^{\infty} dx {x^n-(x+2 \pi
i)^n \over \cosh{x}-\cosh{a}}=-c_1-c_2-c_3-c_4
\eeq
where the $c_i$ are the infinitesimal half circles
surrounding the poles at $\pm a$. The latter ones are
trivially calculated with the final result
\beq
{\rm P} \int\limits_{-\infty}^{\infty} dx {x^n-(x+2 \pi
i)^n \over \cosh{x}-\cosh{a}}&=&{i \pi \over \sinh{a}}
\left ( a^n-(-a)^n \right .\nonumber\\
&+&\left .(2 \pi i + a)^n-(2\pi i-a)^n \right
).\nonumber\\
\label{id}
\eeq
From this identity one now generates the required
integrals $J_n$ by calculating the identity (\ref{id})
for $n+1$. The first two integrals read
\beq
J_0&=&-{2 a \over \sinh{a}}\nonumber\\
J_2&=&-{2 a \over 3 \sinh{a}} (a^2-2 \pi^2).
\eeq
As is obvious from symmetries only integrals $J_n$ with even
numbers of $n$ are nonzero.
The required integral in (\ref{c1}) reads now
\beq
{\rm P} \int\limits_{-\infty}^{\infty} dx {x^2-a^2
\over \cosh{x}-\cosh{a}}&=&J_2-a^2J_0\nonumber\\
&=&{4 a \over 3 \sinh{a}} (\pi^2+a^2).
\label{d1}
\eeq


\newpage
\onecolumn
\begin{table}
\caption{\label{tab1}Comparison between estimates discussed in text and
KB-results. The first two rows are calculated with the situation
of figure \protect\ref{1} and Fig \protect\ref{5}. The last six rows refer
to results with
Fermi distributions as initial conditions.
In the first three of these the density was kept fixed and
the temperature was changed. In the last three
the chemical potential was fixed and the temperature was changed. Temperatures are
given before and after the formation of correlations. The latter
temperatures are higher due to the (negative) correlations. For the correlation energy $E_{\rm corr}$ and the correlation
time $\tau_c$ we compare 3 different expressions with the K/B solution. The numerical values (\protect\ref{eq})
and(\protect\ref{tau}) has been performed by four-dimensional principal value integration. The approximate formulae for
the low temperature limit (\protect\ref{equil}) and (\protect\ref{exact}) are given as well as the high temperature limit
(\protect\ref{vtg}) and (\protect\ref{max1}).}
\begin{tabular}[t]{|l|c|c|c|c|c||c|c|c|c|}
\hline
& $n$ & $\mu(T),\epsilon_f$ & \multicolumn{2}{c|}{$ T$ } & $n
\lambda^3$ &
$ E_{\rm corr}
$ equil.
& $ E_{\rm corr} $ K/B &
$\tau_{c} $ equil.
&
$ \tau_{c}$ K/B
\\
&${\rm fm}^{-3}$&MeV&\multicolumn{2}{c|}{MeV}&&
MeV
&MeV&
${\rm fm/c}$
&${\rm
fm/c}$
\\
&&&init.&fin.&&(\protect\ref{eq}),(\protect\ref{equil}),(\protect\ref{vtg})
&&(\protect\ref{tau}),(\protect\ref{exact}),(\protect\ref{max1})
&
\\ \hline
Fig \protect\ref{1} (case 2) & 0.019 &---, 8.9 &32.7 &36.1 &0.4 &3.0, ---, 7.0
&5   &3.5, ---, 3.5 &3.5
\\ \hline
Fig \protect\ref{5} (case 2) & 0.011 &---, 6.2 &11.6 &13.5 &0.8 &2.1, ---, 5.3
&2.5 &5.6, ---, 4.2 &8.9
\\ \hline
case 4                      & 0.18  &26.8, 39.8&0    &25.1 &6.0 &24.3, 24.2 ,---
&23.4&3.4, 3.5 ,---&3.4
\\ \hline
case 4                      & 0.18  &22.0, 39.8&10   &29.4 &4.7 &24.0, 24.0 ,---
&23.7&3.5, 4.3 ,---&3.5
\\ \hline
case 1                      & 0.18  &-11.7, 39.8&40   &50.2 &2.1 &21.3, --- ,---
&18.3&3.6, --- ,---&4.0
\\ \hline
Fig \protect\ref{ill} (case 4)                      & 0.16  &24.6, 37.1&0    &23.7 &5.9 &22.2, 21.4 ,---
&22.4&3.6, 3.8 ,---&3.4
\\ \hline
Fig \protect\ref{ill} (case 4)                      & 0.18  &22.1, 39.9&10   &29.4 &4.8 &24.0, 24.1 ,---
&23.7&3.5, 4.3 ,---&3.5
\\ \hline
Fig \protect\ref{ill} (case 1)                      & 0.35  &24.9, 62.7&40   &53.6 &3.8 &37.4, 45.5 ,---
&22.4&2.8, 3.8 ,---&4.0
\\ \hline
\end{tabular}
\end{table}

\end{document}